\documentclass[twocolumn, switch]{article} % Method A for two-column formatting

\usepackage{preprint}

\usepackage{amsmath, amsthm, amssymb, amsfonts}
\usepackage{chemformula}

\usepackage[
backend=biber,
style=numeric,
citestyle=phys
]{biblatex}
\usepackage{biblatex}           
\AtEveryBibitem{%
    \clearfield{url}%
    \clearfield{urldate}%
    \clearfield{review}%
    \clearfield{series}%%
    \clearfield{doi}
    \clearfield{issn}
    \clearfield{isbn}
}

\usepackage[colorlinks = true,
            linkcolor = purple,
            urlcolor  = blue,
            citecolor = cyan,
            anchorcolor = black]{hyperref}	% Color links to references, figures, etc.
\usepackage{lineno}		% Line numbers
\usepackage{float}			% Allows for figures within multicol
\usepackage{titlesec}
\titlespacing\section{0pt}{12pt plus 3pt minus 3pt}{1pt plus 1pt minus 1pt}
\titlespacing\subsection{0pt}{10pt plus 3pt minus 3pt}{1pt plus 1pt minus 1pt}
\titlespacing\subsubsection{0pt}{8pt plus 3pt minus 3pt}{1pt plus 1pt minus 1pt}

\title{Reconstruction of 3-D Atomic Distortions from Electron Microscopy with Deep Learning}
\usepackage{xwatermark}
\newwatermark[firstpage,color=gray!90,angle=0,scale=0.18,xpos=0,ypos=130, align=center,width=30in]{This manuscript has been authored by UT-Battelle, LLC under Contract No. DE-AC05-00OR22725 with the U.S. Department of Energy.
The United States Government retains and the publisher, by accepting the article for publication, acknowledges that the United States Government retains a non-exclusive, paid-up, irrevocable, world-wide license to publish or reproduce the published form of this manuscript, or allow others to do so, for United States Government purposes.
The Department of Energy will provide public access to these results of federally sponsored research in accordance with the DOE Public Access Plan (http://energy.gov/downloads/doe-public-access-plan)}

\usepackage{authblk}

\author[1*]{Nouamane Laanait}
\author[2+]{Qian He}
\author[2]{Albina Y. Borisevich}
\affil[1]{Computational Sciences and Engineering Division, Oak Ridge National Laboratory, Oak Ridge, TN, USA}
\affil[2]{Materials Sciences and Technology Division, Oak Ridge National Laboratory, Oak Ridge, TN, USA}
\affil[+]{Present Address: School of Chemistry, Cardiff University, Cardiff, UK}
\affil[*]{laanaitn@ornl.gov}

\begin{document}
\twocolumn[ % Method A for two-column formatting
  \begin{@twocolumnfalse} % Method A for two-column formatting
  
\maketitle
\begin{abstract}
Deep learning has demonstrated superb efficacy in processing imaging data, yet its suitability in solving challenging inverse problems in scientific imaging has not been fully explored. Of immense interest is the determination of local material properties from atomically-resolved imaging, such as electron microscopy, where such information is encoded in subtle and complex data signatures, and whose recovery and interpretation necessitate intensive numerical simulations subject to the requirement of near-perfect knowledge of the experimental setup. We demonstrate that an end-to-end deep learning model can successfully recover 3-dimensional atomic distortions of a variety of oxide perovskite materials from a single 2-dimensional experimental scanning transmission electron (STEM) micrograph, in the process resolving a longstanding question in the recovery of 3-D atomic distortions from STEM experiments. Our results indicate that deep learning is a promising approach to efficiently address unsolved inverse problems in scientific imaging and to underpin novel material investigations at atomic resolution.
\end{abstract}

 % (optional)
\vspace{34pt}

  \end{@twocolumnfalse} % Method A for two-column formatting
]

%%%%%%%%%%%%%%%  Main text   %%%%%%%%%%%%%%%
%  \linenumbers

\section{Introduction}
\begin{figure*}[]
  \centering
  \includegraphics[scale=0.15]{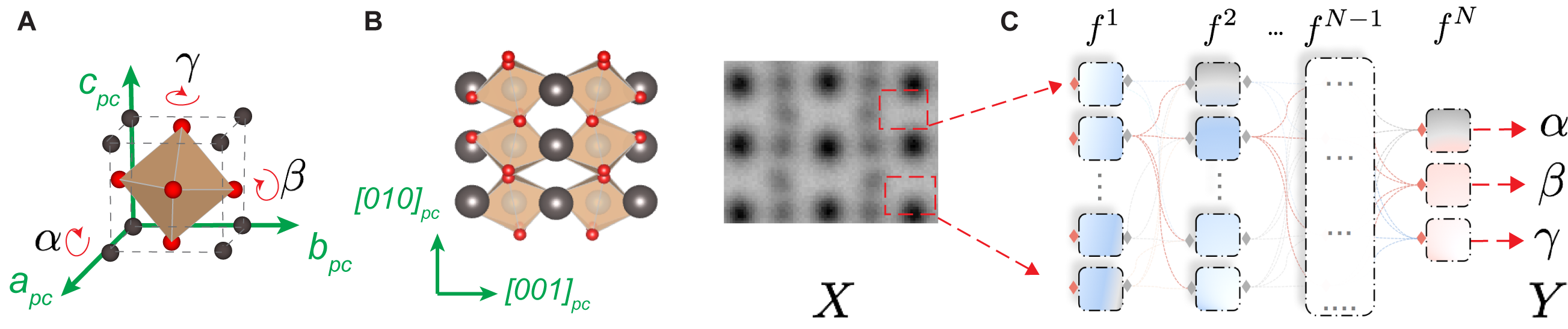}
%   \fbox{\rule[-.5cm]{4cm}{4cm} \rule[-.5cm]{4cm}{0cm}}
  \caption{\textbf{Reconstruction of 3-D octahedral rotations from a single 2-D image with Deep Neural Networks}. (A) Rotations of oxygen octahedra are fully characterized by the magnitude of three angles $(\alpha,\beta,\gamma)$ about the principal crystallographic axes of a perovskite unit-cell ($a_{pc}$, $b_{pc}$, $c_{pc}$; a pseudo-cubic (pc) unit-cell is used throughout and we hereafter omit the pc subscript). (B) Projection along the $[1\bar{1}0]$ axis of a $2\times2$ \ch{CaTiO3} unit-cell (u.c) with tilt signature $a^- b^- c^+$ and the corresponding annular bright-field scanning transmission electron micrograph (ABF-STEM, electron beam propagates along $[1\bar{1}0]$). Octahedral rotations in a perovskite produce anisotropic distortions of the projected oxygen column shape in an ABF-STEM image (box outline). These column shape distortions encode information on the underlying rotation angles $(\alpha,\beta,\gamma)$. Inferring the latter from a single 2-dimensional ABF-STEM image is formulated, here, as a supervised deep learning (DL) problem (C). We seek to construct an optimal mapping $f:X\rightarrow Y$, where X is an ABF image, Y the rotation angles, and $f:= f^N\circ f^{N-1}\circ \dots \circ f^1$ is represented by an N-layer feedforward artificial neural network, where each layer $i$ computes $f^i$.}
  \label{fig:1}
\end{figure*}
Deep learning has made tremendous progress in the past few years\cite{RN1}, and is poised to enable paradigm-changing breakthroughs encompassing the technology industry and the sciences \cite{RN1, RN2}. Of particular interest are deep learning systems geared towards the analysis of imaging data, comprised chiefly of convolutional neural networks\cite{RN3} and variants thereof \cite{RN4, RN5}. These deep artificial neural networks have demonstrated unmatched accuracy and performance at learning to capture the salient and abstract features present in image data and to use such features to perform specific tasks in various fields from clinical pathology \cite{RN6} to astrophysics \cite{RN7, RN8}. Given the minimal assumptions deep learning makes about the nature of the information present in the imaging data \cite{RN9}, it can have broad applicability to imaging studies of materials but has hitherto remained underutilized \cite{RN10}. Here, we demonstrate a deep learning-based solution to a long-standing question in electron microscopy, namely, the ability to infer three-dimensional material structural properties from a single two-dimensional image.

In our study, we target the materials class of complex oxide perovskites whose members host fascinating physical phenomena from correlated electron behavior \cite{RN11, RN12} to quantum magnetism \cite{RN13}, and underpin novel device components such as ferroelectric tunnel junctions \cite{RN14}. One of the defining structural properties of complex oxides perovskites are oxygen octahedral rotations\cite{RN15} (Fig. \ref{fig:1}A). The latter play a central role in the electronic configuration of these materials and consequently their properties via crystal field splitting. Moreover, as symmetry-lowering distortions, oxygen octahedral rotations readily couple to electronic and spin degrees of freedom during phase transitions\cite{RN16}. Research in the field of tilt-driven engineering of electronic and magnetic properties has blossomed (e.g. \cite{RN17, RN18}) and a deeper theoretical understanding of the coupling of tilts to other materials properties has been developed \cite{RN19}. Validation of theoretical predictions of octahedral tilts in strain-engineered heterostructures is primarily carried out against synchrotron surface diffraction measurements, which quantitatively measure the 3-D symmetry and angular magnitudes of octahedral tilts\cite{RN20}, albeit in an ensemble-averaged fashion. In light of the seminal role that local structural states play in influencing the properties and responses of oxides, especially in strain-engineered heterostructures, the need for experimental access to the full local 3-D symmetry and magnitudes of octahedral tilts is indispensable, yet this goal has so far remained elusive.

Direct imaging of abrupt changes in tilts that occur at interfaces or induced by defects has only been possible for a decade \cite{RN21, RN22} due to advances in aberration-corrected electron microscopy. Compared to other structural distortions in perovskites (e.g. strain and polarization), octahedral tilts are more difficult to quantitatively characterize, especially at the local unit cell level, as they are associated with zone-boundary modes and more subtle changes in symmetry. Scanning transmission electron microscopy (STEM) via the annular bright field mode (ABF) can readily resolve atomic columns of light elements such as oxygen. The ABF image, however, is a complex pattern of coherent scattering and interference of electrons through the material \cite{RN23} that is often treated, for simplicity, as a two-dimensional projection of the atomic lattice. While information beyond the projection geometry contributes to the image formation, extracting additional parameters, such as the three-dimensional rotation angles, is a complex inverse problem (Fig. \ref{fig:1}B). Underlying this complexity is the nature of the ABF image contrast which is overwhelmingly dominated by projected information; additionally, atomic columns which produce the most prominent image contrast do not participate in most distortions of the overall crystal structure. \\
Despite these challenges, there has been progress towards extracting 3D local information of octahedral rotations. In particular, an approach was developed for classifying ABF images based on oxygen column shapes \cite{RN24}; it required expert and manual visual inspection of ABF micrographs to identify oxygen columns, thereby isolating the related signal from that of the contrast-dominating cations, followed by a dimensionality-reduction technique (i.e. principal component analysis) to relate the shape of adjacent oxygen columns to the local tilt symmetry. Besides the need for (subjective) input from an expert electron microscope scientist, the main limitation of the previous approach is its inability to associate quantitative 3D octahedral rotation information but for a subset of manually identified ABF oxygen column shapes. \\
In this work, we construct an end-to-end deep learning model to infer from a single ABF-STEM image full and quantitative 3D information of oxygen octahedral rotations (see Fig.\ref{fig:1}C). We find that by training a custom deep convolutional neural network (DCNN) on dynamical electron scattering simulations of perovskite structures, it can extract both symmetry and magnitudes of octahedral rotations from experimental data with unit-cell resolution and sub-degree angular rotations. Our model successfully generalizes what it learned to accurately predict these structural distortions to new material classes it did not see during training, over the entire range of rotation parameter space. The new interpretation of ABF-STEM imaging enabled by a DCNN allows us to quantitatively address the coupling of octahedral distortions, across engineered interfaces of thin-films and superlattices, with atomic resolution, directly from experiments for any oxide system. Furthermore, testing the DCNN on experimental data permits us to identify some of the inherent limitations in extracting 3-D structural information from ABF-STEM experiments.
% \section{Results}
\label{sec:results}
\begin{figure*}[]
\centering
\includegraphics[scale=0.82]{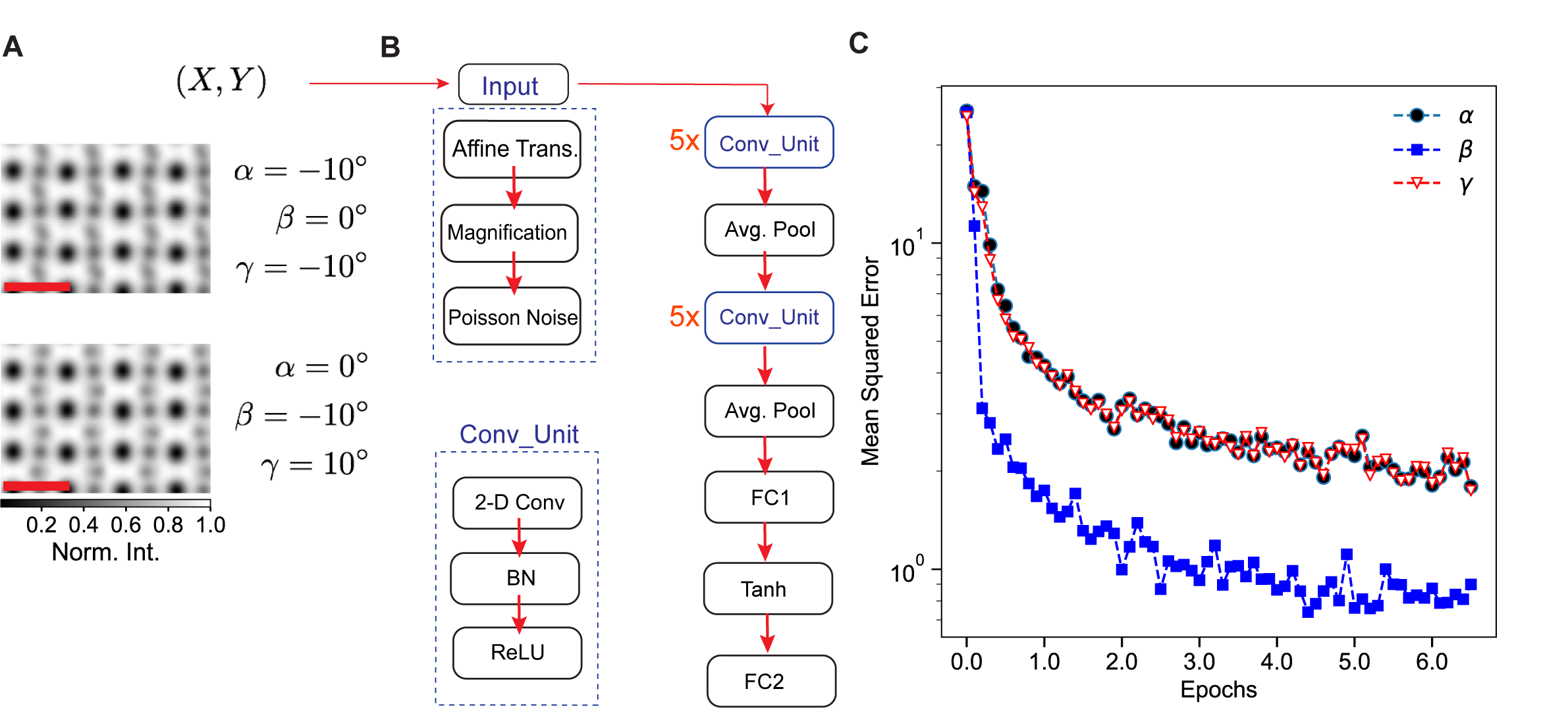}
\caption{\textbf{Training and Validation of a Deep Convolutional Neural Network on Electron Scattering Simulations}. The $(X,Y)$ pairs ($X$: simulated ABF image, $Y$:$(\alpha,\beta,\gamma)$) form the examples on which DCNN training was performed. In total, $7\times10^5$ ABF images of the prototype perovskite \ch{SrTiO3} were generated using multislice simulations and Gaussian processes modeling, encompassing oxygen octahedral rotations spanning a range of $(-10^\circ, 10^\circ)$ in increments of $0.25^\circ$ for each angle. (see Fig.\ref{fig:S1} and Section \ref{section:methods}). Two representative images with different octahedral rotation configurations shown in (A), scale bar is $4\AA$. A 90/10 split of the simulated dataset into training/validation datasets was performed. (B) As DCNN we used a custom 12-layer convolutional architecture Conv Unit: 10 convolutional layers, FC1(2): 2 fully connected layers, BN: Batch Normalization, Avg. Pool: Average Pooling, nonlinear activations are ReLU: rectified linear unit, and $\tanh$) (see Fig. \ref{fig:S2},\ref{fig:S3}). To extend the applicability of the model to data coming from a variety of STEM instruments and settings, we applied to $X$ a sequence of random input image transformations, whose functional forms reflect commonly encountered experimental conditions. (C) The mean-squared error (MSE) per sample between the angles predicted by the DCNN and the true angles from the validation dataset was evaluated concurrently with the training (1 epoch indicates that the DCNN has processed a number of images equivalent to the size of the entire training dataset). The DCNN converges to a per sample validation errors in predicting oxygen octahedral rotation angles on the order of $\approx \pm 1^\circ$.}
\label{fig:2}
\end{figure*}

\paragraph{Structural Refinement in Electron Microscopy with Deep Learning}
In materials with a perovskite structure (\ch{ABO3}, A, B: cations, O: oxygen), rotations of oxygen octahedra (\ch{BO6}) disrupt the perfect alignment of oxygen atoms present in the parent cubic structure and result in oxygen column splitting in projection (Fig. \ref{fig:1}B). While separate oxygen columns can sometimes be visualized, most rotation angles are small ($<10^\circ$, often $<5^\circ$), placing the corresponding oxygen-oxygen separation beyond the resolving power of modern scanning transmission electron microscopes. However, imperfect alignment of oxygen atoms and the associated perturbation of scattering wave-fronts that forms ABF STEM images makes atomic column shapes of oxygen appear distorted (Fig. \ref{fig:1}B). Consequently, the distinct shapes of oxygen columns in an ABF image encode all the information one can access regarding the \ch{BO6} rotations\cite{RN24}. Three angles uniquely determine the rotation pattern, denoted by $(\alpha,\beta,\gamma)$, each indicating a rotation about the principal crystallographic axes $a_{pc}, b_{pc}, c_{pc}$, respectively (we use a pseudo-cubic unit cell throughout and omit the $pc$ subscript hereafter). 
To predict the symmetry and magnitude of a \ch{BO6} octahedron from an ABF image, we need to construct a model that maps column shapes to $(\alpha,\beta,\gamma)$ for every \ch{ABO3} unit cell, whereby the absolute values of these angles give the magnitude, while the symmetry is fixed once the signs of $(\alpha,\beta,\gamma)$ for two neighboring \ch{ABO3} unit-cells are known\cite{RN15}.
The main challenges one encounters in developing a model to extract structural properties from ABF data are: i) nonlinearity in image contrast formation, ii) the ubiquitous effects of dynamical electron scattering, iii) the presence of nontrivial and varying instrumental factors (e.g. lattice distortions due to sample stage drift, beam partial coherence)\cite{RN25}. These three factors make the construction of an analytical model to predict \ch{BO6} rotations directly from an ABF image difficult and consequently such models are currently lacking. 
Here, we introduce a purely computational model in the form of deep convolutional neural networks (DCNN) that overcomes the above limitations. Our approach exploits one of the characteristic properties underlying the modern success of deep neural networks, namely their ability to construct a general mapping $f_{W_i}$,
\begin{equation}
    f_{\mathbf{W}^i }  (\mathbf{X}) = \mathbf{Y},
\end{equation}
where $\mathbf{X}$ is an ABF image, $\mathbf{Y}= (\alpha,\beta,\gamma)$, and $\mathbf{W}^i (i=1,...,N)$ are parameters (or weights) to be learned during training (for each layer $i$ of an $N$-layer neural network, see Fig. \ref{fig:1}C). The ability of a DCNN to approximate the nonlinear relationship between the ABF contrast and the structural distortions is mathematically ensured by the universal approximation theorem, if the network has enough layers and training data \cite{RN26,RN27,RN28}. 
%figure 1
\begin{figure}[]
  \centering
  \includegraphics[scale=0.6]{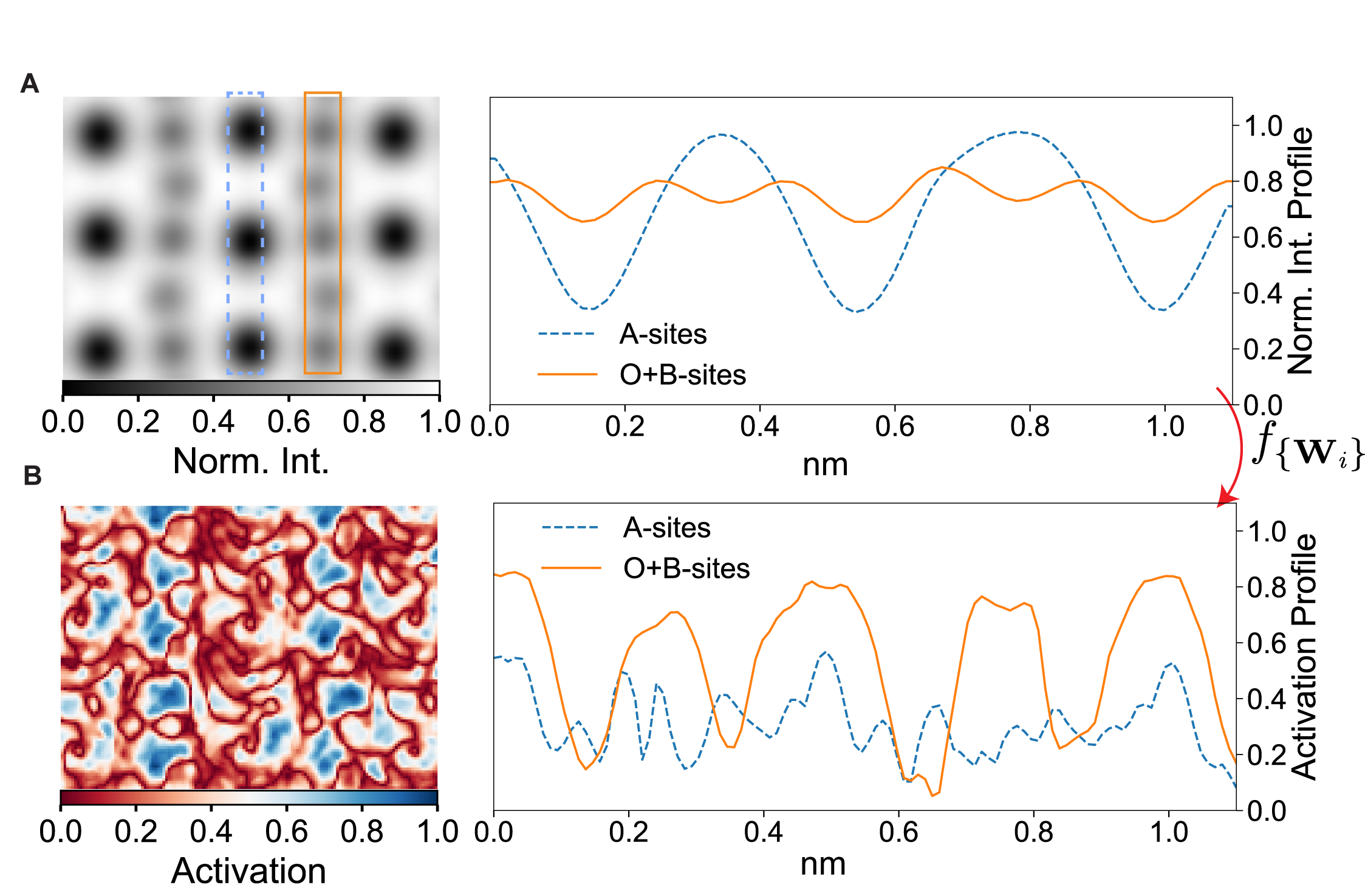}
  \caption{\textbf{Learning to Focus on Oxygen and B-site Atomic Columns}. The forward propagation of an (up-sampled) ABF simulated image in A through the trained neural network produces the activation map in B. This activation map shows that the DCNN is mostly activated by the shape and position of the O-sites as well as the B-sites, while very little “attention” is given to the A-sites. Activation profiles taken along the $a$-axis at the position of A-sites, O- and B-sites (averaged along the $c$-axis inside the box outlines) show the highly localized and strong activations of the latter in comparison to fluctuating and delocalized activations of the former. In essence, the DCNN learned, without any prior information, that to accurately predict oxygen octahedral rotations it should mostly focus on the O- and B-sites (and learned to ignore the strong imaging contrast of the A-sites that is nearly 6 times larger than O or B-cations columns for \ch{SrTiO3}). The normalized absolute value of the activation output is shown and the images in A and B have the same spatial dimensions.}
  \label{fig:3}
\end{figure}

\paragraph{Supervised Training on Electron Scattering Simulations} We took a supervised learning approach to train the DCNN. By using the nearly exact multislice formalism \cite{RN29, RN30} and Gaussian Processes modeling(31), we simulated a dataset of $7\times10^5$ ABF images of the prototype perovskite SrTiO3 oriented in $(110)$ projection. The octahedron rotation angles $(\alpha,\beta,\gamma)$ were modified to span a range of $(-10^\circ, 10^\circ)$ (Fig. \ref{fig:2}A, Fig. \ref{fig:S1}), with the atomic positions of the generated distorted structures subject to the geometric constraint of corner-sharing connectivity \cite{RN32, RN33}. During training, we apply random affine distortions, changes in magnifications, and Poisson noise to each image, to account for commonly encountered instrumental effects (Fig. \ref{fig:2}B). In essence, by training and validating the DCNN on multislice simulated ABF images, the model will incorporate any and all effects of dynamical electron scattering. Moreover, by accounting for instrumental artifacts in the simulated ABF images, the DCNN can learn image features that are more relevant to imaging data encountered in experiments.
The DCNN architecture used is a custom 12-layer convolutional neural network that was trained with an adaptive stochastic gradient-based optimization technique to minimize the Huber loss function between the predicted angles and the correct angles associated with each training image. We find that the validation mean-squared error (MSE) between predicted and true angles converges to $\Delta \alpha,\Delta \gamma \approx \pm 1.4^\circ$ and $\Delta \beta \approx \pm 1^\circ$ after a few data epochs (Fig. \ref{fig:2}C). Interestingly, we find that the validation errors for $\alpha,\gamma$ are correlated, especially during latter stages of the learning phase. Such correlation is a possible indication that the DCNN is learning shared ABF features in its identification of $\alpha,\gamma$ rotations. From a materials perspective, the crystal projection used in our simulations (and later STEM experiments) does indeed couple the 2-D image formation from these 2 \ch{BO6} rotations. We show below that this statistical correlation does not appear to affect the predictions of the DCNN on experimental images adversely. Finally, we note that our model’s angular errors compare favorably to those obtained from structure refinement of spatially-averaged electron diffraction measurements ( $\pm 0.72^\circ$ for the \ch{BO6} rotations of the low-temperature phase of \ch{SrTiO3})\cite{RN34}.

\paragraph{learning to Ignore}
Arguably, the main drawback of any DCNN model is the difficulty in interpreting, in a human-accessible form, what it learned and how it makes its predictions, irrespective of how accurate or inaccurate they may be \cite{RN28, RN35}. We find that in this particular application one can arrive at a qualitative physical interpretation of what the DCNN learned during training by propagating an ABF image (from the validation set) through the network to obtain an activation map (Fig. \ref{fig:3}A, B). First, note that the bright-field contrast for most \ch{ABO3} perovskites will be primarily dominated by the A-cations; for \ch{SrTiO3} they produce contrast approximately five times larger than the contrast from O or B-site columns (line profile in Fig. \ref{fig:3}A). Second, an activation map indicates what parts of the image give a stronger response from the DCNN and is qualitatively interpreted as what parts of the image the model focuses on. Remarkably, despite this significant imbalance in contrast between different atomic columns, we find that the trained model learned to focus on the oxygen and B-cation columns with well-localized activations, while it almost entirely ignores the presence of A-cations (line profile in Fig. \ref{fig:3}B). These results are entirely consistent with our physical understanding of octahedral distortions in perovskites involving only displacements of oxygen and B-cations to an excellent approximation. Such physical understanding was not given as input to the neural network but was learned solely during training. Moreover, it is interesting that the DCNN is equally activated by the oxygen and B-cations, despite the latter’s minute changes from one octahedral distortion configuration to next (compare the two simulated ABF images at different sets of angles in Fig. \ref{fig:2}A). The latter observation is one of the hallmarks of deep learning, whereby a deep neural network will learn the features it needs directly from the data\cite{RN1}, without input from a user, to accomplish the task at hand (see Fig. \ref{fig:S3}).

\begin{figure}[H]
  \centering
  \includegraphics[scale=0.2]{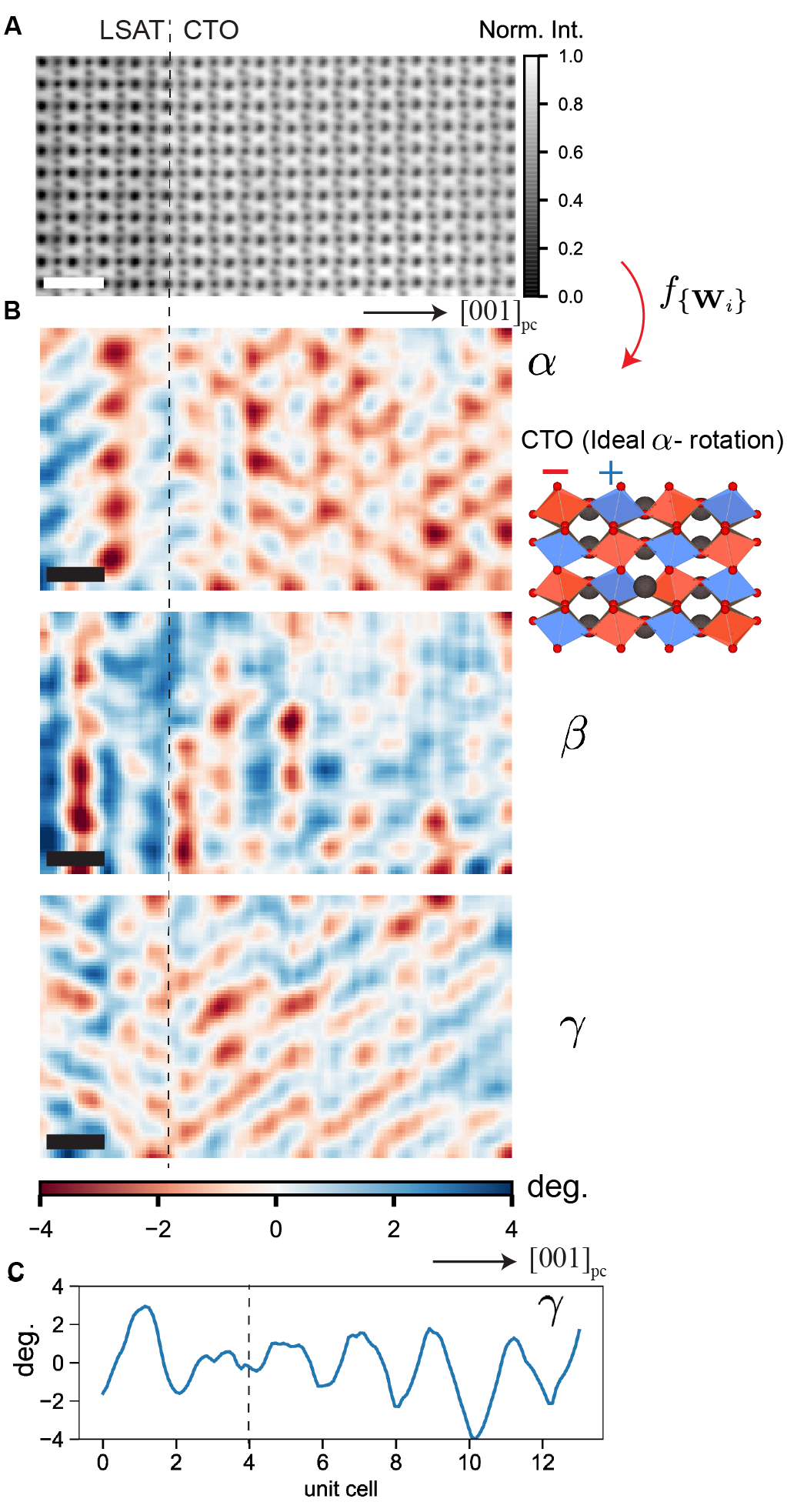}
  \caption{\textbf{Inference of Oxygen Octahedral Rotations Symmetry and Magnitudes from Experiment}. (A) Experimental ABF STEM micrograph of a \ch{CaTiO3} (CTO) epitaxial thin film on a single crystal substrate \ch{(LaAlO3) 0.3 (Sr2AlTaO6) 0.7} (LSAT). (B) Predictions of \ch{BO6} octahedral rotation angles spatial distributions with the trained model $f_{W_i }$  match the tilt pattern of CTO $(a^- b^+ c^-)$, whereby rotations about the $a-$ and $c$-axis $(\alpha,\gamma)$ alternate from positive to negative. The ideal CTO tilt pattern is illustrated for $\alpha$-rotation. (C) The well-known periodicity of oxygen octahedral rotations in CTO (2 unit cells) is also reproduced by the DCNN as shown in the profile of $\gamma$-rotations along the c-axis (see Fig. \ref{fig:S3} for all line profiles). DCNN predictions of $\beta$-angles match the expected symmetry of CTO, with notable exceptions near the interface with LSAT, where we found negative rotations. In fact, the magnitudes of both in-plane angles $(\alpha,\beta)$ in the first 2 CTO unit cells from the interface fluctuate substantially, an observation that is consistent with the elastic constraints imposed by crystalline epitaxy. The latter also induces distortions of oxygen octahedra at the surface of LSAT causing deviations from its expected bulk symmetry ($a^0 b^0 c^0$, no rotations). The vertical dashed line in all subfigures outlines the approximate position of the interface. The scale bars are 8\AA (~ 2 unit cells). During inference with the DCNN, minimal preprocessing of the experimental data via global intensity scaling was performed (see Section \ref{section:methods} and Fig. \ref{fig:S4} for inference procedure).}
  \label{fig:4}
\end{figure}

\paragraph{Transferring Knowledge Learned from Simulations to Experiments}
To determine if the DCNN trained exclusively on simulated data of \ch{SrTiO3} is nevertheless effective in extracting structural properties from experimental data and generalizes to other \ch{ABO3} perovskites, we tested its performance on ABF-STEM experimental images of an epitaxial thin-film of \ch{CaTiO3} (CTO) on a single crystal substrate \ch{(LaAlO3) 0.3 (Sr2AlTaO6) 0.7} (LSAT) (Fig. \ref{fig:4}A). 

In thin-film form, \ch{CaTiO3} has an octahedral tilt symmetry of $a^- b^+ c^-$ (24), indicating that the $\alpha$ and $\gamma$ angles of two neighboring unit cells have opposite signs, while their $\beta$ rotations are of the same sign. We found that our trained model correctly predicts the octahedral rotation symmetry for \ch{CaTiO3} far from its interface with LSAT (> 2 u.c) with a spatial resolution at the level of a single unit cell. The spatial distributions of $\alpha$ and $\gamma$ display alternating rotations with the well-known unit-cell doubling periodicity (Fig. \ref{fig:4}B, C). Near the interface with LSAT, we see strong modulations in the sign of the octahedron rotation about the b-axis indicating that the tilt pattern of those first 4 CTO unit cells is a mixture of $a^- b^- c^-$ and $a^- b^+ c^-$. The presence of a different rotation pattern at an epitaxial interface, especially the rotation about an in-plane crystallographic axis (i.e., orthogonal to the growth direction [001]), is fully consistent with the mechanical boundary conditions imposed on the CTO by LSAT via the misfit strain. 

In the case of LSAT, the DCNN predictions do not match its bulk pattern ($a^0 a^0 a^0$, no rotations) near the interface. Instead, we see large magnitudes in $\alpha$ and $\beta$ ($\approx 4^\circ$) corresponding to octahedral rotations about the in-plane axes, $a$ and $b$, respectively. Inspection by eye confirms that the LSAT oxygen column shapes are in fact distorted, pointing to the potential presence of rotations. In principle, at a symmetry-changing interface in an epitaxial heterostructure, such as CTO/LSAT, the substrate (LSAT) and certainly the film (CTO) could exhibit structural distortions in a finite transition region that are distinct from the bulk crystal structure.\cite{RN19, RN36} Moreover, the observed 2-fold larger magnitudes of in-plane rotations (i.e., $\alpha$ and $\beta$) than the out-of-plane angle are consistent with the elastic constraints imposed by epitaxy and imply that the substrate interfacial unit-cells partially “inherit” the rotation pattern of CTO. The presence of these rotations and their asymmetry was further corroborated by analysis of additional ABF-STEM experimental data (Fig. \ref{fig:S6}). Due to the limited imaging field of view in the experimental data, we were unable to unambiguously quantify the extent of this transition region and consequently confirm that LSAT reverts to its bulk structure far away from the interface.\\

%figure 2

\begin{figure*}[!ht]
  \centering
  \includegraphics[scale=0.22]{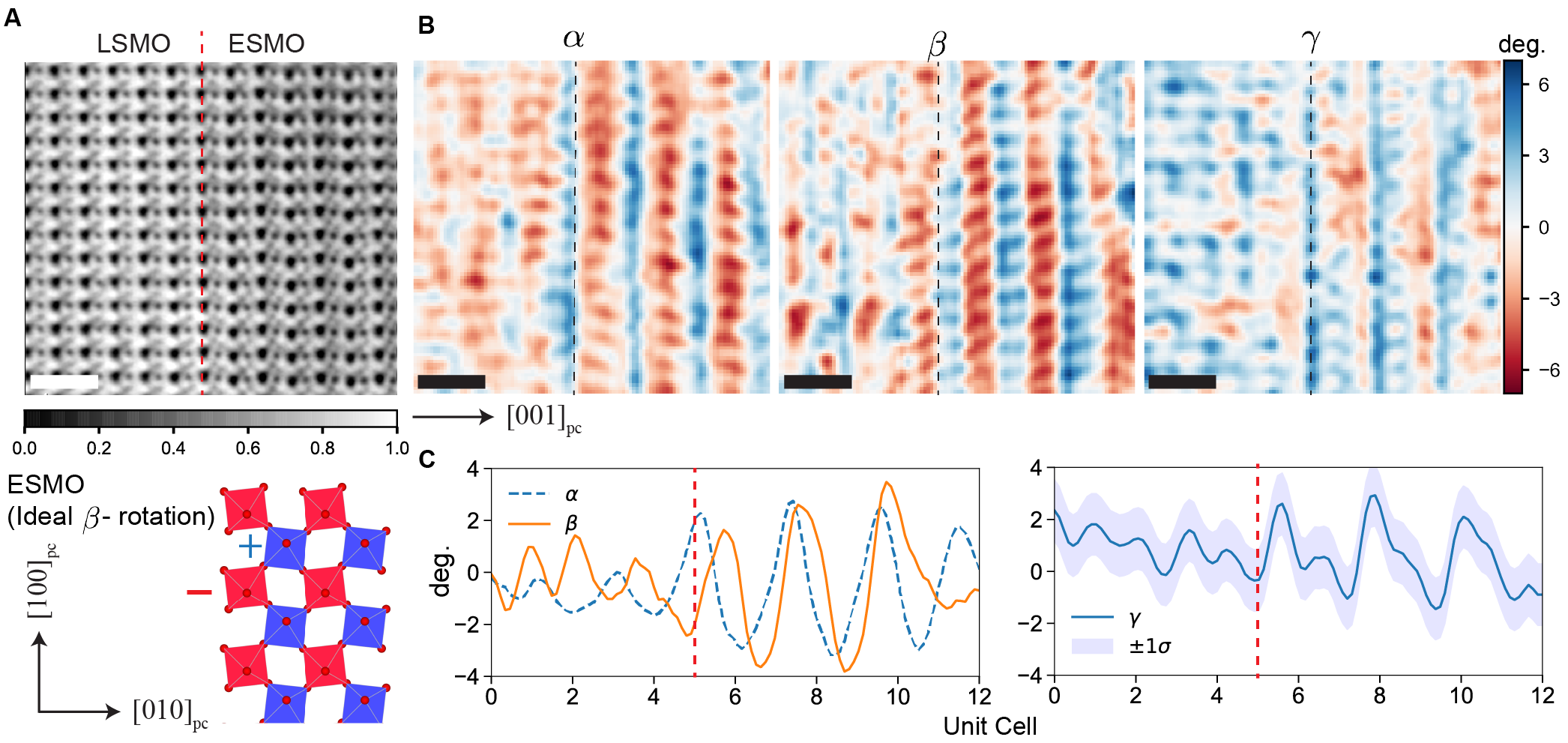}
  \caption{\textbf{Deep Learning Generalization to New Material Classes}. (A) Experimental Annular Bright Field STEM micrograph of \ch{La0.7Sr0.3MnO3} / \ch{Eu0.7Sr0.3MnO3} (LSMO/ESMO) superlattices. (B) Prediction of \ch{BO6} octahedral rotations by the DCNN trained on \ch{SrTiO3} electron simulations. Despite having no prior knowledge of ESMO’s orthorhombic perovskite structure (inset in A), our model correctly predicts the oxygen octahedral symmetries and magnitudes of the material, whereby alternating positive/negative in-plane angles $(\alpha,\beta)$ in ESMO arise out of the underlying symmetry class ($a^- a^- c^+$), with a small position-averaged $\gamma$ rotations of $\approx 0.36^\circ$ (C) Rotation profiles of the spatial distributions in (A) were averaged along [100]. For LSMO, with a rotation pattern of $a^- a^- c^-$, weak alternating rotation sense for $(\alpha,\beta)$ are found but are within the uncertainty bounds of the DCNN derived from the validation dataset. Our model does not accurately capture the value of rotations about the $c$-axis in LSAT due to pronounced shape distortions of all atomic columns, likely arising from a misalignment between the electron beam axis and the zone-axis of LSMO. The vertical dashed line in all subfigures outlines the approximate position of the interface. Scale bar is 10\AA (~ 2 unit cells). The confidence bands in C represent $\pm1\sigma$ derived from the per sample validation mean-squared error and assume that the model is a statistically unbiased estimator.}
  \label{fig:5}
\end{figure*}

\paragraph{Generalization to the Wider Family of Oxides}
To rigorously test the generalization of our DCNN model in accurately predicting octahedral rotations of the wider family of complex oxides, we tested it on ABF-STEM experimental images of \ch{La 0.7 Sr 0.3 MnO3 / Eu 0.7 Sr 0.3 MnO3 (LSMO/ESMO)} superlattices (Fig. \ref{fig:5}A). The crystal structure of ESMO, in particular, represents a substantial departure from the structural configurations our deep learning model was trained on, as it belongs to a different class of orthorhombic-distorted perovskites (i.e., \ch{GdFeO3}-type structure). In addition to the octahedral tilting the DCNN encountered during training, the structure of ESMO contains distortions of the A-cation sub-lattice which were not part of the training set, and in bulk form, is found in the $a^- a^- c^+$ octahedral rotation pattern\cite{RN36}. \\
Our model predictions are in excellent agreement with the tilt pattern of ESMO (Fig. \ref{fig:5}B), especially for $\alpha$ and $\beta$, whose absolute values are identical to less than $1^\circ$ as required by the octahedral rotation symmetry class (Fig. \ref{fig:5}C). In the case of $\gamma$, we find the absence of distinct alternating out of phase rotations (ruling out a $c^-$ pattern) and the presence of prominent fluctuations in the spatial distributions of octahedral rotations that average out to $\approx 0.4^\circ$ or, equivalently an $a^- a^- c^+$. Note, however, that the mean-squared prediction error MSE for the out-of-plane rotation is $\Delta \gamma \approx \pm 1.4^\circ$(per sample in the validation set, see Fig. \ref{fig:2}C). Consequently, while this deep neural network could not distinguish between a $c^+$ or $c^0$ pattern with high confidence from this particular ABF-STEM data set, it can effectively generalize to other oxide material classes, not seen during training, and correctly and quantitatively infer their structural properties from raw experimental data. \\
In the case of LSMO, with a bulk octahedral rotation pattern of $a^- a^- c^-$, alternating rotation sense in the $\alpha$ and $\beta$ channels of the DCNN analysis are discernable, in agreement with the bulk structure. In the case of out-of-plane rotations, our model does not reproduce the bulk symmetry in predicting a $c^+$ pattern with a spatially-averaged rotation angle of $\gamma= 1.2^\circ \pm 1.4 ^\circ$. A closer inspection of the LSMO micrograph indicates the presence of strong shape anisotropies affecting all atomic columns, most likely arising out of a zone-axis misalignment during the experimental acquisition. The incorrect out-of-plane octahedral symmetry predictions for LSMO and the absence of sharp contrast in the rotation phase of $\alpha$ and $\beta$ (relative to ESMO and CTO, Fig. \ref{fig:5}C) indicate that the robustness of the deep learning model to pronounced misalignments could be further improved.

\section{Discussion \& Conclusion}
Previous studies have noted that oxygen column shapes in ABF STEM data contain information about the 3-D symmetry of perovskites, but quantitative information on the latter was deemed irretrievable but for a small subset of tilt systems.\cite{RN24} The high accuracy and reliability displayed by a deep learning model in assigning tilt systems and extracting tilt angles across the entire span of rotation symmetries and magnitudes suggests otherwise. Essentially, the presented results point to the presence of interpretable signatures in ABF STEM data that have evaded the trained eye of human experts and previous analyses, which can be learnt by a deep convolutional neural network and used to quantify 3-D structural distortions of oxides, directly and unassisted, from experimental data. 

The central aspect of the success of this deep convolutional neural network is its capability to focus solely on those image features relevant to octahedral distortions and learning to ignore, altogether, far more intense yet irrelevant image features, such as ABF contrast of A-cations. This physical understanding that the neural network learned unaided allows it to successfully predict the underlying structural properties, and such understanding simply cannot be encoded in standard electron microscopy analysis routines without the reliance on persistent input from a human expert with a requisite knowledge of oxides. 

Remarkably, “learning to ignore” also underlies the robustness and wide applicability of this deep learning approach: while trained solely on simulated data of simple tilted perovskites with a \ch{SrTiO3} composition, it is nevertheless able to seamlessly generalize to experimental data of entirely different chemical compositions of \ch{CaTiO3}, \ch{La0.7 Sr0.3MnO3}, and \ch{(LaAlO3)0.3(Sr2AlTaO6)0.7}, and even to markedly different structures such as ESMO’s distorted perovskite. Such generalization is not achievable with library-based or look-up tables approaches, which fail to provide meaningful results for cases not present in the library. In fact, it is this ability to generalize to unseen cases that is at the heart of current interest in deep learning. An important aspect of the generalization of the DCNN model of crucial practical importance lies in its ability to accurately infer structural information from experimental data which inherently contains residual microscope aberrations, despite being trained on aberration-free simulated data. 

In contrast to applications of deep neural networks in other fields, such as medical imaging \cite{RN37}, established knowledge of the structure of materials at the atomic level allows a qualitative interpretation of the CNN's activations patterns to analyze its current performance in various settings, and to shed light on how it may be extended to new ones. For instance, by focusing on the relative positions and shapes of the oxygen and B-cation columns (Fig. \ref{fig:3}), the model is less sensitive to changes in contrast, for instance, due to different atomic occupations in the A- or B-sites (e.g., LSAT) or distortions in the positions of the A-cations (e.g., ESMO). However, since the network’s activations are highly localized at these atomic sites, in the presence of severe distortions of atomic column shapes due to small misalignments from the optical axis its predictions will deteriorate less if those same activations were less localized. 

Besides zone-axis misalignments, relative misorientations between substrate and thin-film are also prevalent in experimental STEM studies of epitaxial heterostructures and could limit the applicability of this model with highly-localized activations. It is worth exploring if the model can learn to recognize such common experimental factors and overcome these inherent limitations of STEM once re-trained, using techniques of transfer learning, on new datasets that incorporates optical aberrations, beam tilt variations, and structural imperfections. Another avenue for improving the interpretive abilities of our neural network lies in using more sophisticated simulation approaches to electron scattering, to take into account beam partial coherence and inelastic scattering\cite{RN38}. Given the current approach’s universal generality and good performance, one can credibly expect that incorporation of all additional factors will only improve the precision and accuracy of predicted octahedral distortions. It remains to be determined, however, if all of these parameters can be feasibly incorporated into a single neural network.

Previous experimental studies have probed coupling of octahedral rotations across epitaxial interfaces, finding indirect and qualitative evidence, in the form of 2-d projections, of the substrate’s interfacial unit cells “inheriting” a tilt pattern from the film \cite{RN19, RN24, RN39}. As demonstrated for the CTO/LSAT system, deep learning provides the first fully quantitative evidence of this effect with unit-cell spatial resolutions and excellent angular sensitivity. Moreover, the 3-D configurations extracted by the DCNN can finally enable a direct comparison between STEM measurements and theoretical predictions of 3-D octahedral rotations. Finally, the well-known computational efficacy of deep neural networks can allow for analysis of gargantuan STEM data volumes and henceforth collection of statistically robust structural properties for meaningful comparisons with other structural probes such as X-rays \cite{RN20}.  
The ability of the end-to-end deep learning system, generated in this study, to adequately represent the complex relationships between annular bright field contrast and the underlying three-dimensional atomic crystal structure of a material, suggests that the potential of deep learning extends beyond the automation of routine image analysis tasks. And it can be harnessed to address challenging inverse problems in imaging \cite{RN40} with spatially coherent probes \cite{RN41}, thereby powering the discovery of hitherto inaccessible local materials properties.

\section*{Acknowledgments}
\footnotesize

\paragraph{Acknowledgments}
We would like to acknowledge Steve May group (Drexel University) for providing samples of the ESMO/LSMO superlattices, and Michael Biegalski (Oak Ridge National Lab) for providing samples of LSAT/CTO. This work used resources of the Oak Ridge Leadership Computing Facility (OLCF) at Oak Ridge National Laboratory, which is supported by the Office of Science of the Department of Energy under Contract DE-AC05-00OR22725. AB acknowledges support from the Materials Science and Engineering Division of the US DOE Office of Science, Basic Energy Sciences.

\paragraph{Author Contributions}
NL and AB designed the research. NL performed the modeling and computation. QH performed the scattering simulations. All authors discussed the results. NL and AB wrote the paper.

\paragraph{Competing Interests} 
The authors have no competing nor conflicts of interests.

\paragraph{Materials and Correspondence}
Reasonable requests for data and computer code used should be addressed to N. Laanait (laanaitn@ornl.gov).

\paragraph{Data Availability and Computer Code} 
Dataset used to train and validate the deep learning model will be published and accessible through the figshare data repository (a placeholder DOI: 10.6084/m9.figshare.6826880 has been reserved for the dataset pending acceptance of the manuscript). Computer Code used in this work (deep learning and Gaussian Processes Modeling) will be posted at https://github.com/nlaanait/DeepLearnSTEM.git\\

%%%%%%%%%%%% Supplementary Methods %%%%%%%%%%%%
\footnotesize
\section{Methods}
\label{section:methods}
\subsection{STEM Experiments and Simulations}

\paragraph{Experiments}
Scanning transmission electron microscopy data were collected using a Nion UltraSTEM microscope operated at 200 kV Annular Bright field detector with angular acceptance range of 15-30 mrad range was used for recording ABF images. Sample preparation details for all materials can be found in \cite{RN24}.

\paragraph{Simulations}
Structure models for \ch{SrTiO3} composition with oxyge  octahedral rotations angles spanning a range of $(-10^\circ, 10^\circ)$ in increments of $5^\circ$ for each $(\alpha,\beta,\gamma)$ angle were generated using the software \textit{POTATO}, which incorporates geometrical constraints related to corner sharing connectivity of the octahedra\cite{RN32, RN33}. These models were then rotated and translated to form a rectangular cuboid with axes of $[110]$, $[001]$ and $[1\bar{1}0]$ and size $2\sqrt{2}\times4\times80\sqrt{2}$ pseudo-cubic unit cell. Multislice simulations using Kirkland frozen phonon codes were then used to compute ABF (15-30 mrad) images for the inner $\sqrt{2}\times2$ span of pseudocubic unit cells (to avoid edge effects) and sample thicknesses from 2 to 42 nm. A 200 kV aberration-free probe with 30 mrad probe-forming aperture was used for the calculation, and the results were averaged over 15 random thermal configurations generated for the temperature of 300K.

\paragraph{Gaussian Processes Machine Learning} 
Deep learning models require vast amounts of training data ranging from $10^5$ to $10^7$ samples (i.e. ABF images). Performing such a large number of multislice simulations is computationally prohibitive for most materials and beyond the computational budget of most researchers. Instead of such a brute-force simulation approach, we employed Gaussian Processes modeling (GP), a nonparametric and multidimensional Bayesian modeling technique \cite{RN31}. With GP, we used our multislice simulations of ABF image contrast at angular increments of $5^\circ$ to find a surrogate model, that is computationally cheaper and highly accurate in predicting ABF images at finer angular increments (see Fig. \ref{fig:S1}). Formally, let us denote our multislice simulations of the ABF contrast by $C_0$, and the angular rotations by a rotation vector $\bf{\theta}=(\alpha,\beta,\gamma)$. The GP surrogate model of the ABF contrast, $C_{GP} (\theta)$, is given by,
\begin{equation}
    C_{GP} (\mathbf{\theta})=\phi(\mathbf{\theta}^T )\mathbf{w},\mathbf{w} \sim \mathcal{N}(m,\Sigma)
\label{eq:gp}
\end{equation}
where $\phi$ is a kernel-based mapping of the angular vector into feature space, $\mathbf{w}$ are (unknown) parameters to be determined from $C_0$, and $\mathcal{N}$ is a multivariate normal distribution with a mean (vector) $m$, and covariance matrix $\Sigma$. The covariance matrix, $\Sigma=\sigma ^2  k(\theta,\theta')$  is fully determined by our choice of kernel $k$ and the variance $\sigma$ is estimated by maximum likelihood with our multislice simulated $C_0$. In our GP modeling we used a Matern kernel, given by
\begin{equation}
    k(x)=  \frac{2^{(1-\nu)}}{\Gamma(\nu)} \bigl(\frac{\sqrt{2\nu x}}{l}\bigr)^\nu K_\nu \bigl(\frac{\sqrt{2\nu x}}{l}\bigr)
\end{equation}

where $K_\nu$ is a modified Bessel function, $\Gamma$ is the gamma function, and $l$ is a hyper-parameter of the Gaussian Process (we also tested a radial basis function kernel but found that it gave larger mean-squared errors than the Matern kernel, see Fig. \ref{fig:S1}). During the fitting stage of the GP model, we used $\nu=3/2, l=1$, and conjugate gradient descent minimization of the log-likelihood. We repeated the minimization procedure 20 times with random $l$ - parameter initialization, and the GP model, $C(\theta)$, with the minimal marginal log-likelihood was chosen. After optimization, we found that the GP modeling of the ABF image contrast, $C_{GP}$ (defined in Eq.\ref{eq:gp}) gave a very good approximation to the multislice simulated contrast $C_0$, with mean-squared errors on the order of $10^{-4}$. From the trained GP model, we sampled ABF images spanning a range of $(-11.25^\circ, 11.25^\circ)$ in increments of $0.25^\circ$ for each $(\alpha,\beta,\gamma)$, for a total of 729,000 ABF images. We found that for absolute angular values larger than $11.25^\circ$, the GP posterior covariance increased and as such the model cannot be reliably used to predict ABF contrast for octahedral rotation states outside of the range $(-11.25^\circ, 11.25^\circ)$. The numerical implementation of Gaussian Processes in the \textit{scikit-learn} library was used throughout this work\cite{RN42}.

\subsection{Deep Learning}

\paragraph{Training/Validation Data}
The simulated dataset consisting of 729,000 ABF images and $(\alpha,\beta,\gamma)$ labels was partitioned into training and validation sets using a 90/10 split.  
During training we pre-processed each training batch before feeding into the network using a combination of global affine distortions, global scaling, and noise sampled from a Poisson distribution. These transformations reflect some of the commonly encountered experimental conditions \cite{RN25}. For instance, global affine distortions approximate the geometric distortions of an image due to sample stage drift, while Poisson noise reflects the counting statistics in a scattering process. Unlike scanning distortions and counting statistics, uncontrolled changes in magnification rarely occur in practice on the same instrument, under the same settings, and are typically known with good accuracy and precision. 

The training data included such magnification changes largely to facilitate inference from experimental images, and to aid our deep neural network to generalize to small changes in magnification $(\approx \pm 5\%)$ that can enable it to draw from data pools coming from different microscopes and settings (see Subsection \ref{subsection:model_prediction}). The functional form of all of the above transformations is fixed during training, but their defining parameters (e.g. horizontal and vertical shears, 2 rotation angles and 2 translations for an affine distortion) were randomly sampled from a uniform distribution for each $(X,Y)$ in a training batch. Moreover, we also used the common practice of random crops of the images in the training batch. The physical dimensions of the image crops were taken as a $2\times2$ $(101)$ projected unit cells of \ch{SrTiO3} $(\approx 8 \AA \times 11 \AA)$ as to not break the structural symmetry of some octahedral rotation patterns.

\paragraph{Deep neural network architecture} 
Our deep convolutional neural network (DCNN) is a custom 12 layers architecture (Conv Unit: 10 convolutional layers, FC1(2): 2 fully connected layers). Batch normalization (BN) was used before the nonlinear activation via ReLU (rectified linear unit) and after the 2-D convolution layers (2-D Conv). A downsizing of the input at different stages of the DCNN was performed with average pooling layers (Avg. Pool). The $\tanh$ nonlinear activation is used prior to the linear output layer (FC2). Figure \ref{fig:S2} contains a list of all parameters (kernel sizes, strides, etc…) that fully fix the architecture of our DCNN.

Our large training/validation image database allowed us to not rely on techniques of transfer learning, and therefore we trained our model from scratch (i.e. random initialization). In light of the markedly different intensity distributions and data channels of STEM images relative to images of natural scenes and objects which are predominantly used in designing new DCNN architectures (e.g. natural image databases such as ImageNet), we chose to design our own custom DCNN architecture to achieve optimal results in the task of reconstructing octahedral rotations from a single 2-D image. 

In our DCNN architecture design, we used commonly used neural network layers (2-D convolutions, batch normalization, etc…) \cite{RN4} and followed the well-established practices of: (i) composing layers such that the height and width of the features decreases with DCNN depth, while increasing the number of these features, and (ii) increasing the representation capacity of the DCNN via depth to improve generalization \cite{RN9}.

During our DCNN architecture design, we found that the use of max pooling layers consistently produced poor validation accuracies. Such max pooling layers are predominantly used in state of the art DCNN models trained on natural image databases. We attribute the unsuitability of max pooling to the sparsity of ABF STEM data, the pronounced absence of well pronounced edges, and the strong localization of information (at the atomic scale). The preceding characteristics are almost never satisfied in natural image databases, and further motivates our use of a custom DCNN architecture. As shown in Fig. \ref{fig:S3}, the filters learned by the first convolutional layer of our DCNN model are devoid of “edge” filters, which are commonly found in DCNN models trained on natural image databases, supporting our observations regarding the difference between scientific images such as STEM ABF and natural image databases. 

An interesting future direction, will be to fully quantify the performance of state of the art DCNN models such as ResNet(4) and DenseNet(5) on our simulated and experimental STEM data, using our custom model as reference, due to its proven capability to accurately infer materials properties, and to generalize to new materials and new imaging conditions.

\paragraph{Model Training} 
The deep learning library \textit{Tensorflow} (v1.4) was used to implement and optimize the DCNN \cite{RN43}. Training was performed on an NVIDIA DGX-1 system ($8\times$ Tesla P100 GPU) using data parallelism. 
We trained our DCNN model using an adaptive stochastic gradient descent algorithm (ADAM optimizer, $\beta_1=0.9, \beta_2=0.999$, with a staircase learning rate decay policy) to minimize the loss function $\mathcal{L}$ given by  
\begin{equation}
\label{eq:loss}
    \mathcal{L}(Y,Y') = \mathcal{L}_{Huber}(Y,Y') + \epsilon \Sigma_i ||W_i||^2 ,	
\end{equation}

where $\mathcal{L}_{Huber}$ is the Huber loss evaluated on the true labels $Y$ (i.e. $\alpha,\beta,\gamma$) and the predicted labels $Y'$ (described below). The second term on the right-hand side of Eq. \ref{eq:loss} is an $L_2$ regularization term with coefficient $\epsilon = 10^{-4}$, and the sum index $i$ runs over the layers of the neural network, with $\mathbf{W}_i$ denoting a vector of weights of layer $i$. The Huber loss is a commonly used loss function in robust regression to reduce the effect of outliers (relative to a mean-squared error loss) and is by given by
\begin{equation}
    \mathcal{L}_{Huber} (Y,Y')= 
    \begin{cases}
        \frac{1}{2} (Y-Y')^2,& \text{if } |Y-Y'|\leq \delta\\ 
        \delta |Y-Y'|- \frac{1}{2} \delta^2, & \text{otherwise}
    \end{cases}                    
\end{equation}

We used an initial “cutoff” value $\delta=25^\circ$ during training for each octahedral rotation angle. The value of $\delta$ was decayed using the same schedule and decay rate as the learning rate (see caption to Fig. \ref{fig:S2}).

\paragraph{Model Validation} 
Validation of the model was quantified by computing the mean-squared error between the predicted angles and the true angles over the entire validation dataset (= 72,000 ABF images and labels). The validation mean-squared error value we report in Fig. \ref{fig:2}C (the main text) is averaged over all predictions of the DCNN (for each angle) and is repeatedly evaluated from a saved copy of the DCNN model throughout training. No ABF image augmentation is performed before validation

\paragraph{Model Prediction from Experimental Data}
\label{subsection:model_prediction}
Prediction of the octahedral rotation spatial maps from experimental STEM ABF images reported in the main text is performed using sliding windows. Each sliding window extracts an image patch, whose size spans approximately $2\times2$ $(101)$ projected unit cells, from an experimental STEM ABF image and used as input into the DCNN to predict the angular rotations (see Fig. \ref{fig:S4}). This procedure is repeated as the sliding window is scanned across the entire experimental image with a stride size of 1 pixel (equal in height and width). The predictions of the model are independent of the stride size due to the intrinsic translation invariance property of convolutional neural networks and the additional fact that during training the model sees randomly extracted image patches (i.e. random crops) extracted from the same ABF image and are all associated with the same octahedral rotation state (see Model Training Subsection). Moreover, the model predictions are not sensitive to the exact physical size of the image patch, since the DCNN was trained on ABF STEM data at different magnifications. Finally, a moving average kernel whose size corresponds to $1\times1$ unit-cell is applied to the model predictions (oversampled by the sliding windows) to produce octahedral rotations with unit-cell spatial resolutions, as reported in Figs. \ref{fig:4},\ref{fig:5} in the main text.

\paragraph{Activation Map} 
The activation map reported in Fig.\ref{fig:3} in the main text was obtained by transforming the trained fully-connected layers (FC1, FC2) into 2-D convolutional layers with a $1\times1$ kernel size, and forward propagating an up-sampled ($\times8$) ABF image, randomly chosen from the validation set, through the entire DCNN. The output of this transformed DCNN is an image with three channels (formerly corresponding to each angle). We reported the absolute mean value over the three channels with the intensities scaled in the range $[0,1]$.

%%%%%%%% References %%%%%%%%%%%
% \normalsize
% \bibliography{CTO_DL.bib}
\printbibliography
%%%%%%%%%%%%  Supplementary Figures  %%%%%%%%%%%%
\clearpage
\section{Supplementary Figures}
\renewcommand\thefigure{S\arabic{figure}}
\setcounter{figure}{0}

\begin{figure}[!ht]
  \centering
  \includegraphics[scale=1.0]{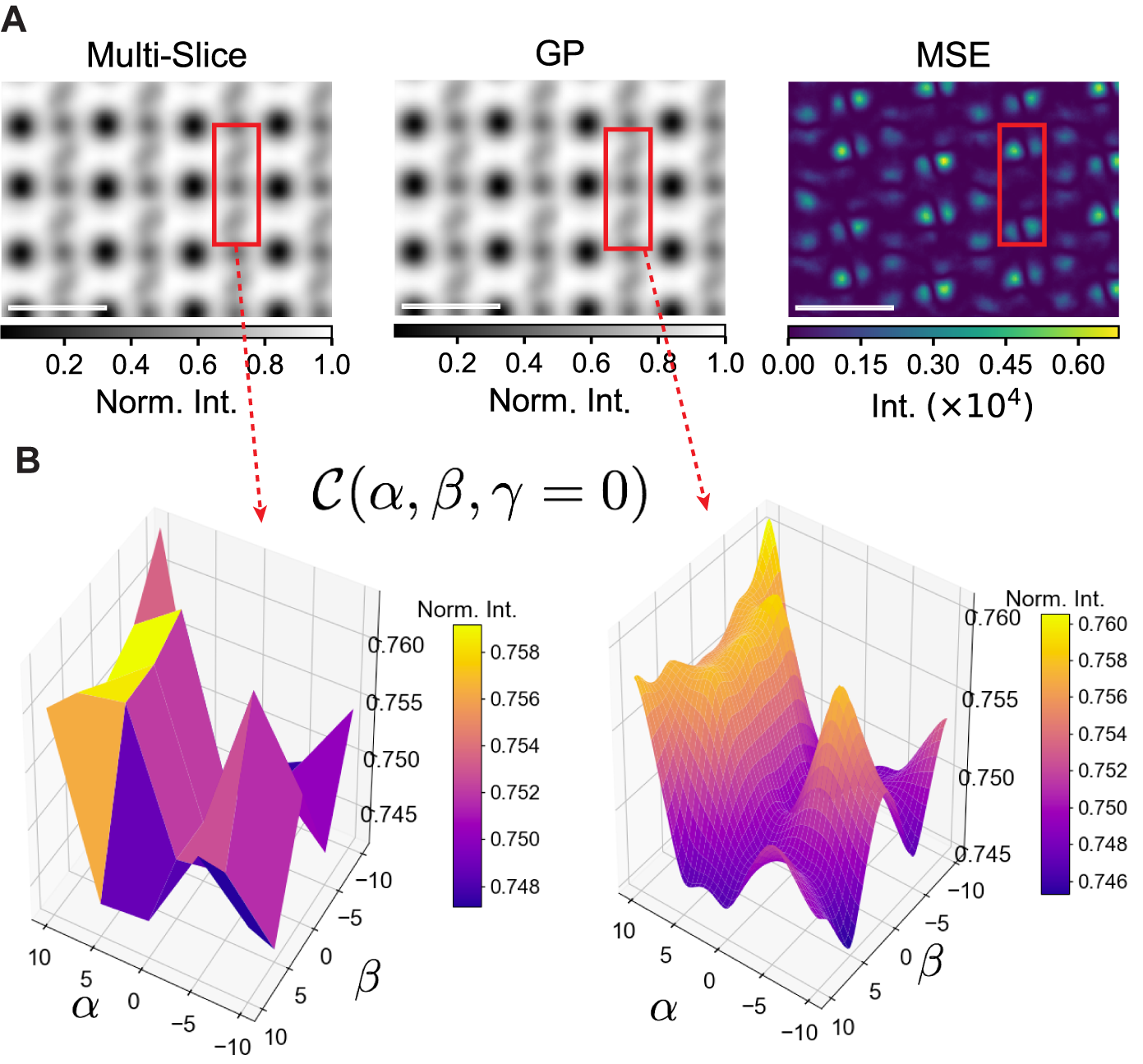}
  \caption{\textbf{Gaussian Processes Modeling of the Dynamical Contrast Function of Oxygen Octahedral Rotations}. (A) Annular bright field image at fixed $(\alpha_o,\beta_o,\gamma_o)$ predicted by Multi-Slice (not used for training) can be used to estimate accuracy of Gaussian Processes (GP) modeling image at same angular values. Mean squared error (MSE) map between the Multi-Slice image and the GP modeled, reveal that GP modeling errors are on the order of $\pm 10^{-2}$ of the normalized intensity and are maximal at spatial locations of oxygen columns (box outline). The observed behavior is consistent with the physical consideration that column shape and intensities of A-cations do not vary as function of $(\alpha,\beta,\gamma)$ in the angular range of consideration $[-10^\circ, 10^\circ]$. (B) The contrast function $C_0 (\alpha,\beta,\gamma=0)$, in steps of $5^\circ$, predicted by multi-slice calculation in a spatially averaged region spanning the atomic columns in the box outline, and the GP modeled $C_{GP} (\alpha,\beta,\gamma=0)$ in angular steps of $0.25^\circ$. During the fitting stage of the Gaussian Processes, we used a Matern kernel (radial basis function kernel was also tested but gave MSE on the order of $10^{-2}$).}
  \label{fig:S1}
\end{figure}

\begin{figure}[!ht]
  \centering
  \includegraphics[scale=1.0]{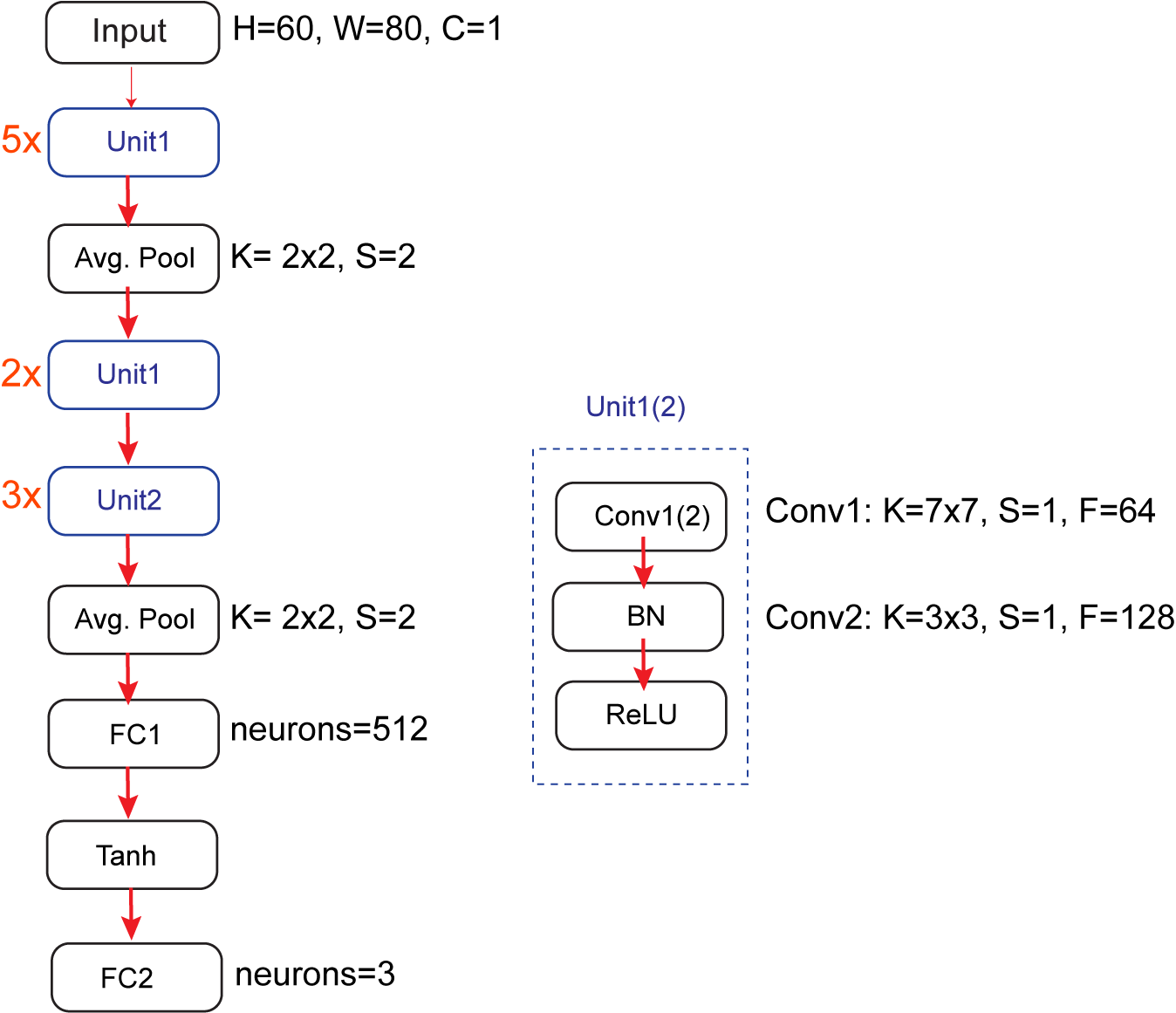}
  \caption{\textbf{The architecture of the Deep Learning Model}. The input is a batch of images, where each image has height H, width W, and channels C. Unit1(2) are composed of a 2-D convolutional layer, with kernel size K, stride S, and number of filters F. The relevant parameters for the average pooling layers (Avg. Pool) and fully-connected layers (FC1(2)) are also indicated. A padding mode of `SAME` is used in the convolutional and pooling layers. BN: Batch normalization. The hyper-parameters used were: batch size = 32, initial learning rate=1e-3, and a learning rate decay = 0.5 was applied every 2 epochs of data. For batch normalization, $\epsilon 10^{-3}$, and an exponential moving average decay of 0.9 to accumulate mean and variance statistics during training for use in inference. All layer weights were initialized using He initialization.}
  \label{fig:S2}
\end{figure}

\begin{figure}[!ht]
  \centering
  \includegraphics[scale=0.4]{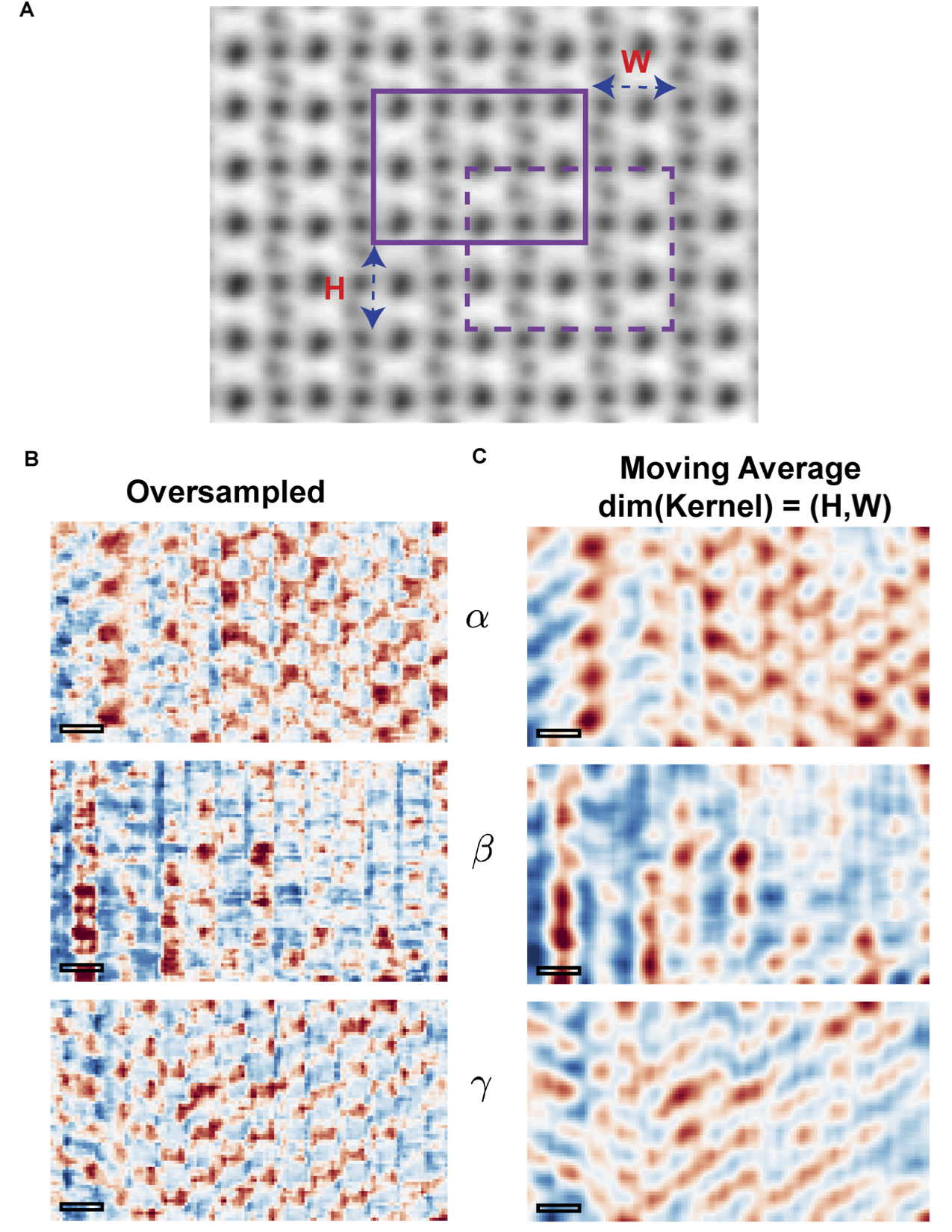}
  \caption{\textbf{Convolution Kernels Learned by the First Layer}. In various other deep learning applications, where a DCNN is trained on natural scenes and subjects (e.g. cats), one finds that the filters learned by the first layer of the network contain a large proportion of edge filters. Instead, we find that the learned filters in this case are nearly devoid of edge filters, an observation that is explainable by the absence of well pronounced edges in scanning transmission electron micrographs at the atomic scale.}
  \label{fig:S3}
\end{figure}

\begin{figure}[!ht]
  \centering
  \includegraphics[scale=0.17]{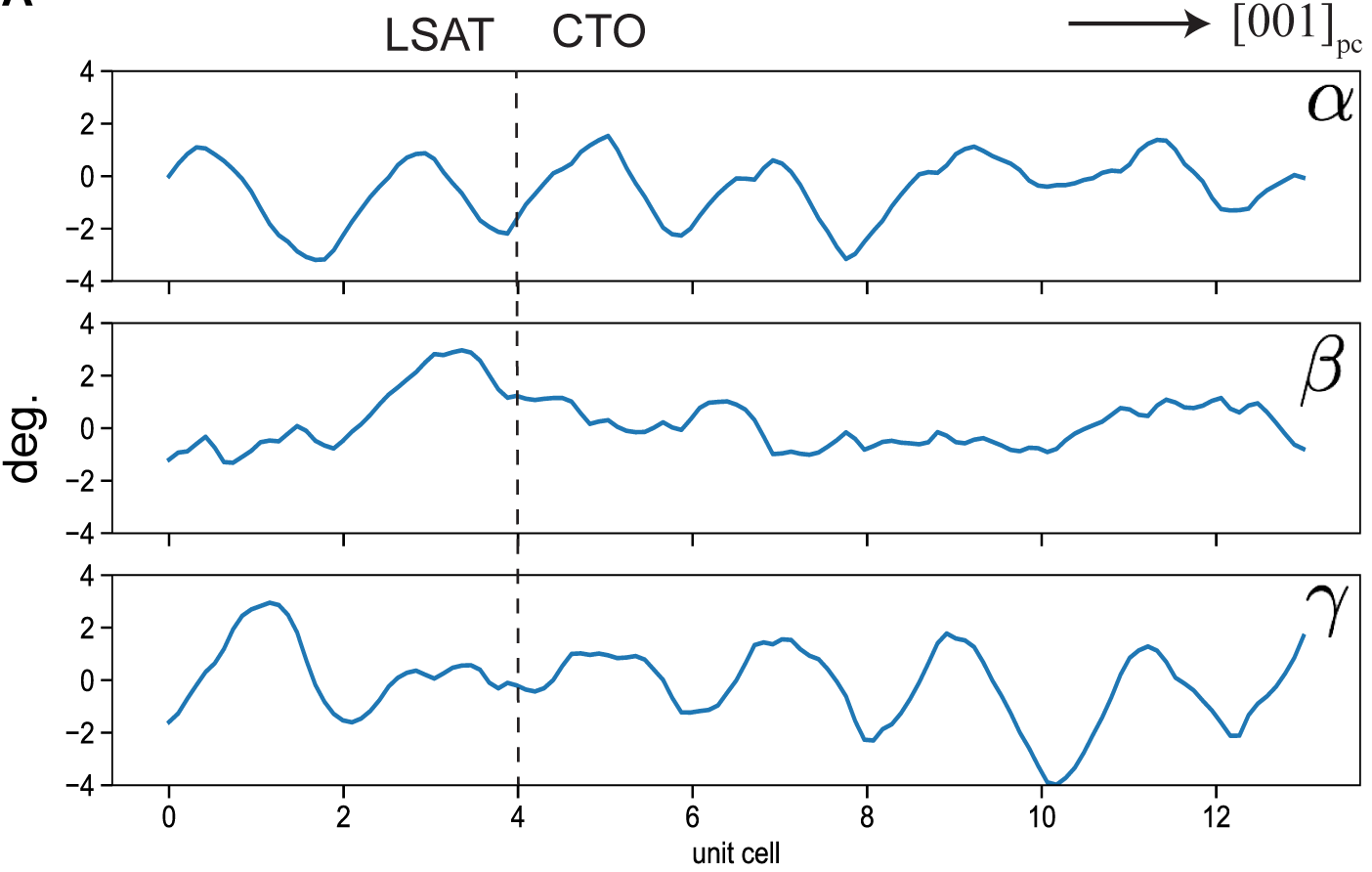}
  \caption{\textbf{Prediction of Spatially-Resolved Octahedral Rotations from Experimental Data}. (A) Two sliding windows from an experimental ABF STEM image of \ch{CaTiO3}. At each sliding window location an image patch is extracted and input into the DCNN model to predict $(\alpha,\beta,\gamma)$. The resultant angular spatial maps are shown in (B). A moving average kernel of size $(H,W)$ corresponding to a $1\times1$ projected unit cell is applied to the oversampled predictions in (B) to obtain spatial maps of octahedral rotations with unit-cell spatial resolutions.}
  \label{fig:S4}
\end{figure}

\begin{figure}[!ht]
  \centering
  \includegraphics[scale=0.8]{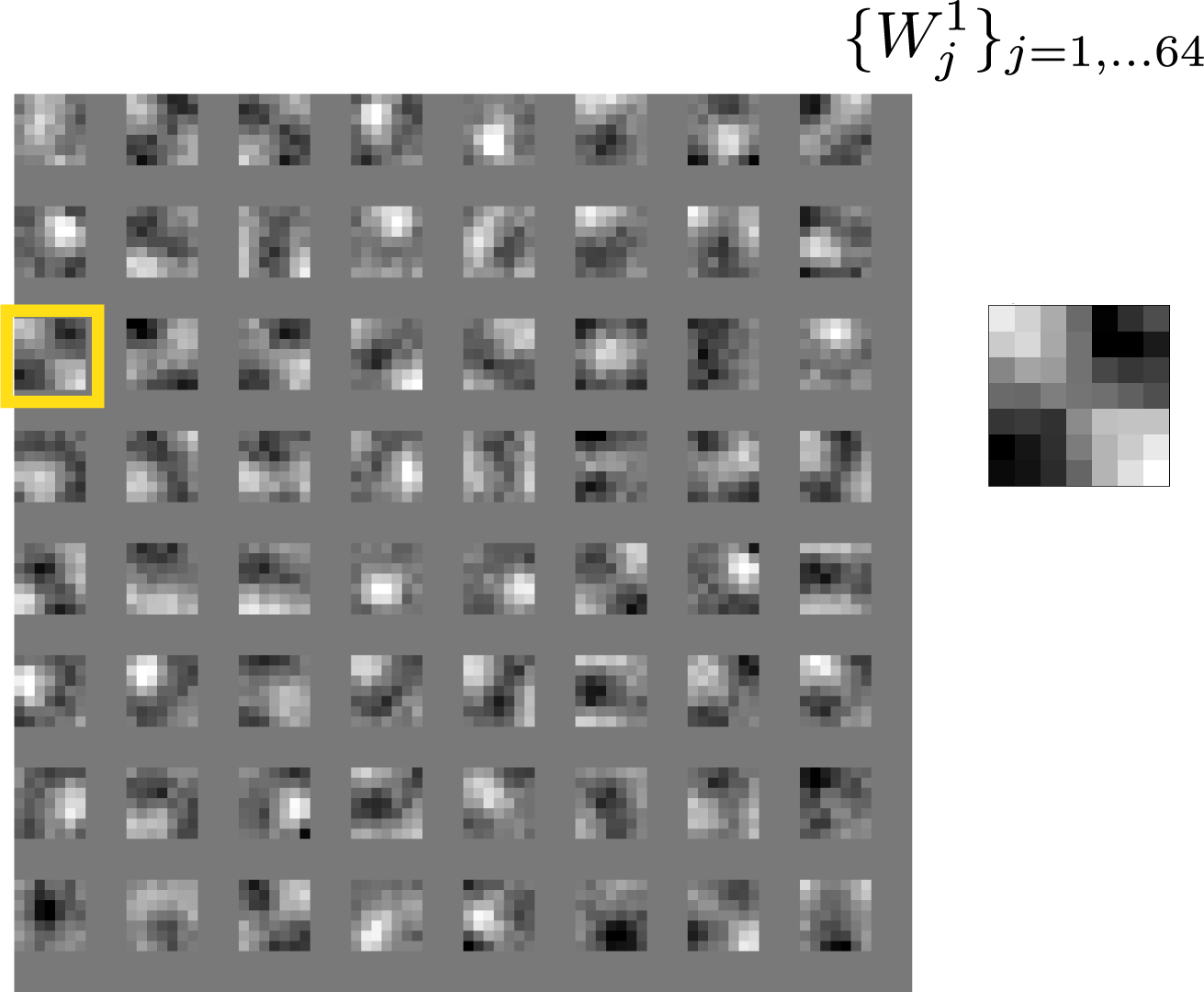}
  \caption{\textbf{Line Profiles of Octahedral Rotations}. Complete line profiles across spatial distributions of octahedral rotations predicted by the DCNN for the CTO/LSAT material shown in Fig. \ref{fig:4} (B). Dashed lines indicate the approximate positions of the interface.}
  \label{fig:S5}
\end{figure}

\begin{figure*}[!ht]
  \centering
  \includegraphics[scale=0.275]{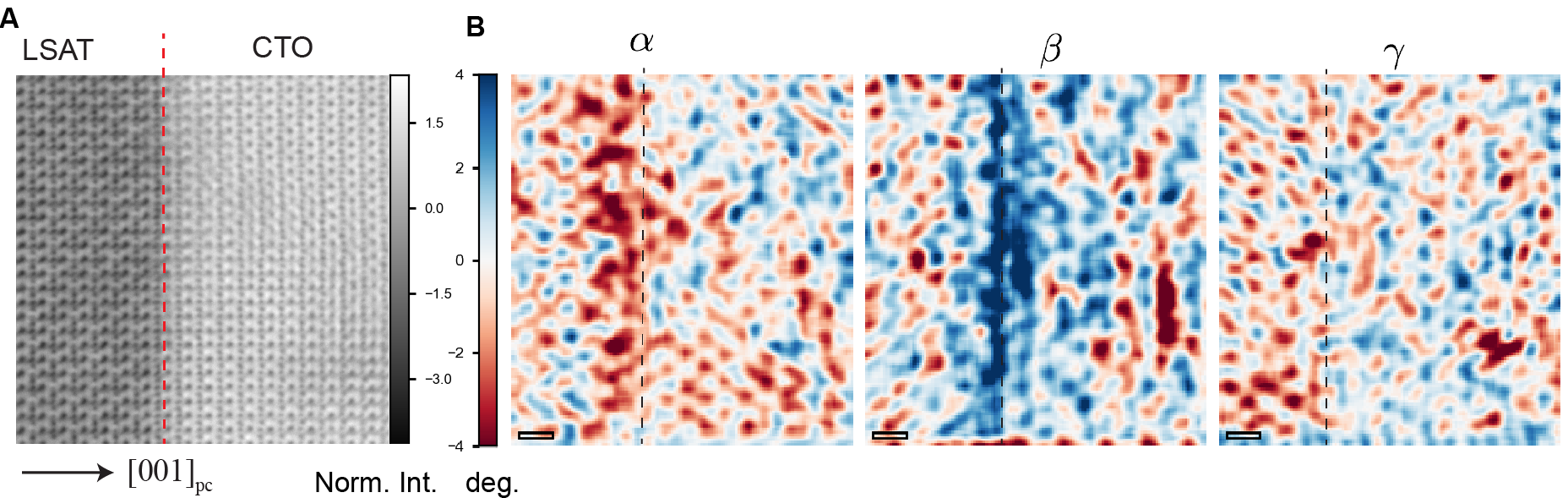}
  \caption{\textbf{Deep Learning Predictions on Additional Experimental data}. The presence of misorientations and the diffuse interface is visually apparent in (A), and is consistent with the DCNN predictions in (B), as manifested by (i) the lack of clear unit-cell doubling (checkerboard pattern in the upperhalf of the CTO image) and (ii) large angular magnitudes in the octahedral rotations that are most affected by crystalline epitaxy $(\alpha,\beta)$, well-localized at the interface.}
  \label{fig:S6}
\end{figure*}

%%%%%%%%%%%%%%%%   End   %%%%%%%%%%%%%%%%
\end{document}